\pdfoutput=1

\documentclass[11pt]{article}

\usepackage[final]{acl}

\usepackage{booktabs}
\usepackage{multirow}
\usepackage{caption}
\usepackage{array}
\usepackage{setspace}
\usepackage{adjustbox}
\usepackage{times} 
\usepackage{amsmath}
\usepackage{afterpage}
\usepackage{latexsym}
\usepackage{subcaption}
\usepackage{tabularx}
\usepackage{xcolor}
\usepackage[normalem]{ulem}
\usepackage{lipsum}
\usepackage{amssymb}
\usepackage{longtable}
\usepackage{lscape}
\usepackage{amsfonts}

\usepackage[T1]{fontenc}
\usepackage[utf8]{inputenc}

\usepackage{microtype}
\usepackage{stfloats}
\usepackage{inconsolata} 

\usepackage{graphicx}
\usepackage{cleveref}

\usepackage{hyperref}

\useunder{\uline}{\ul}{}

\title{Use of Retrieval-Augmented Large Language Model Agent for Long-Form COVID-19 Fact-Checking}

\author{
Jingyi Huang\textsuperscript{$\spadesuit$}, Yuyi Yang\textsuperscript{$\heartsuit$}, Mengmeng Ji\textsuperscript{$\heartsuit$}, Charles Alba\textsuperscript{$\heartsuit$}, Sheng Zhang\textsuperscript{$\spadesuit$}, Ruopeng An\textsuperscript{$\diamondsuit$}\\[0.5em]
\textsuperscript{$\diamondsuit$}New York University, USA\\
\textsuperscript{$\heartsuit$}Washington University in St. Louis, USA\\
\textsuperscript{$\spadesuit$}Shanghai University of Sports, China\\[0.5em]
\parbox{\linewidth}{\centering\texttt{
\href{mailto:jh9925@nyu.edu}{\textcolor{black}{jh9925@nyu.edu}},
\{\href{mailto:y.yuyi@wustl.edu}{\textcolor{black}{y.yuyi}},
\href{j.mengmeng@wustl.edu}{\textcolor{black}{j.mengmeng}},
\href{alba@wustl.edu}{\textcolor{black}{alba}}\}@wustl.edu, 
\href{mailto:zhsheng1@126.com}{\textcolor{black}{zhsheng1@126.com}},\\
\href{mailto:ra4605@nyu.edu}{\textcolor{black}{ra4605@nyu.edu}}}}}

\begin{document}
\maketitle
\begin{abstract}
The COVID-19 infodemic calls for scalable fact-checking solutions that handle long-form misinformation with accuracy and reliability. This study presents SAFE (system for accurate fact extraction and evaluation), an agent system that combines large language models with retrieval-augmented generation (RAG) to improve automated fact-checking of long-form COVID-19 misinformation. SAFE includes two agents—one for claim extraction and another for claim verification using LOTR-RAG, which leverages a 130,000-document COVID-19 research corpus. An enhanced variant, SAFE (LOTR-RAG + SRAG), incorporates Self-RAG to refine retrieval via query rewriting. We evaluated both systems on 50 fake news articles (2–17 pages) containing 246 annotated claims (M = 4.922, SD = 3.186), labeled as true (14.1\%), partly true (14.4\%), false (27.0\%), partly false (2.2\%), and misleading (21.0\%) by public health professionals. SAFE systems significantly outperformed baseline LLMs in all metrics (p < 0.001). For consistency (0–1 scale), SAFE (LOTR-RAG) scored 0.629, exceeding both SAFE (+SRAG) (0.577) and the baseline (0.279). In subjective evaluations (0–4 Likert scale), SAFE (LOTR-RAG) also achieved the highest average ratings in usefulness (3.640), clearness (3.800), and authenticity (3.526). Adding SRAG slightly reduced overall performance, except for a minor gain in clearness. SAFE demonstrates robust improvements in long-form COVID-19 fact-checking by addressing LLM limitations in consistency and explainability. The core LOTR-RAG design proved more effective than its SRAG-augmented variant, offering a strong foundation for scalable misinformation mitigation.

\end{abstract}

\section{Introduction}
\subsection{Background}

The COVID-19 pandemic was accompanied by an “infodemic” marked by the rapid spread of misinformation and disinformation, which significantly undermined public health efforts\cite{thelancet2020infodemic}. Misinformation involves the unintentional spread of false or misleading content  \cite{borges2022infodemics} while disinformation refers to the deliberate dissemination of falsehoods to deceive \cite {wang2022understanding}. Both phenomena contribute to the dissemination of inaccurate information, with studies showing that up to 46.6\% of widely shared posts during the pandemic contained inaccurate claims, contributing to high-risk health behaviors\cite{yang2021battleground, wilhelm2023measuring}. 

Long-form fake news posed a particular challenge, as complex narratives made it more persuasive and harder to debunk \cite{andrew2024understanding, rodrigues2024social, leite2025cross}. Many such articles spread falsehoods about illness severity, vaccine safety, treatment efficacy, and the virus’s origins\cite{atehortua2021covid, reuters2024misleading}, underscoring the urgent need for reliable fact-checking tools capable of addressing detailed and deceptive long-form content to combat the “infodemic”.

Conventional COVID-19 fact-checking methods face major challenges when applied to long-form content. Relying on human experts or traditional machine learning (ML) models is often time-consuming and impractical for texts with complex narratives \cite{kolluri2022covid}. Large language models (LLMs), such as ChatGPT, offer a potential solution, but they are prone to hallucinations—generating plausible yet factually incorrect information not grounded in real-world knowledge or input data \cite{augenstein2024factuality, ji2023survey, wei2024long}. These errors stem from LLMs’ reliance on limited training data, lack of real-world understanding, and their probabilistic generation process. Moreover, LLMs struggle with long-form inputs due to token limits and architectural constraints \cite{tang2024minicheck}. One notable issue is the “lost in the middle”, where models tend to miss or misinterpret content located in the middle of a long document, despite performing better at the beginning or end \cite{baker2024lost}. 

Retrieval-augmented generation (RAG) technique enables LLMs to leverage external information when generating responses, significantly reducing the risk of producing inaccurate content \cite{lewis2020retrieval}. By integrating external data and incorporating it into the model’s context, RAG helps address the issue of hallucinations in LLMs \cite{gao2023retrieval}. This approach provides a way to keep LLMs current without the need for expensive retraining or fine-tuning, making it a practical choice for ensuring reliable and up-to-date performance in various tasks \cite{cai2024forag}. 

Recent studies empirically confirm the effectiveness of RAG in fact-checking. \citeauthor{momii2024rag}  achieved a 0.387 score on AVeriTeC—far surpassing the 0.11 baseline—using question generation and prompt refinement. \citeauthor{sevgili2024uhh} ranked 6th of 23 with a vector-based RAG approach. \citeauthor{russo2024face} further show that LLM-based retrievers and zero-shot prompts yield high verdict faithfulness in realistic settings, underscoring RAG’s robustness across diverse inputs.

A multi-agent system is a collaborative architecture where specialized agents work together to handle complex tasks more effectively than a single LLM \cite{yang2024llm}, particularly in long-form tasks. In this study, a preprocessing agent segments lengthy texts and extracts COVID-19-related claims, which are then passed to downstream agents for fact-checking using an advanced RAG framework. We refer to this integrated approach as the System for Accurate Fact Extraction and Evaluation (SAFE).

\subsection{Objective and Contributions}

This study proposes the SAFE system aims to improve the automatic fact-checking of long-form content related to COVID-19. By integrating an agentic chain architecture and a contextual dataset based on approximately 130,000 peer-reviewed COVID-19 publications, we evaluated the performance of the SAFE system configurations built on LLM agents and RAG-enhanced models. This study makes four key contributions. First, we develop an agentic RAG-based system that reduces both time and computational costs compared to conventional fact-checking methods. Second, the SAFE system improves the reliability of LLMs by grounding fact-checking in peer-reviewed papers, reducing hallucinations and external factual inaccuracies. Third, the integration of agentic preprocessing and advanced RAG techniques addresses the limitations of LLMs in handling long-form input and output, particularly the “lost in the middle” issue \cite{baker2024lost}. Finally, the proposed SAFE system addresses the challenge of fact-checking lengthy COVID-19 articles by ensuring contextual coherence and factual grounding, paving the way for real-world deployment in infodemic combat.

\section{Methods}

\subsection{Workflow}

\begin{figure}[ht!]
    \centering
    \includegraphics[width=\linewidth]{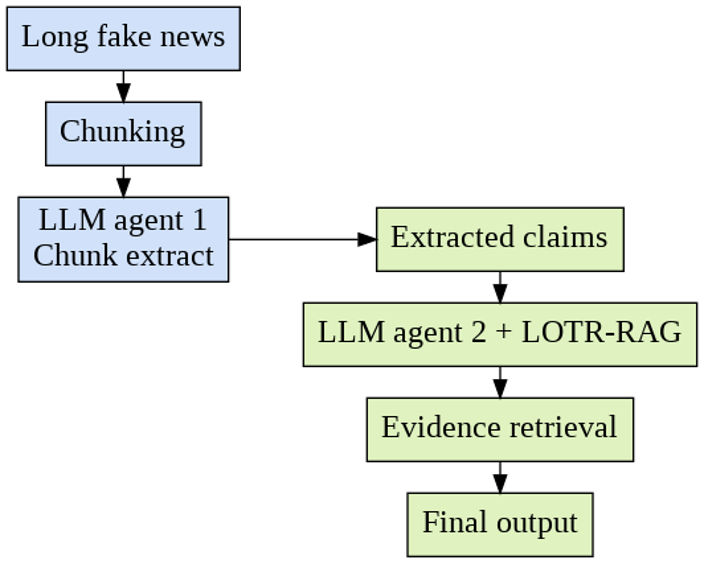}
    \caption{Figure 1. Workflow of the SAFE System}
        \label{fig: Workflow}
\end{figure}

Figure \ref{fig: Workflow} illustrates the workflow of the SAFE system. The process begins by dividing long fake news into overlapping chunks (2,000 tokens with a 200-token overlap) to maintain semantic continuity and fit within LLM input limits. The first LLM agent extracts COVID-19  claims from each segment, which are then passed to a second agent for fact-checking. This agent employs Lord of the Retrievers (LOTR), an advanced RAG framework \cite{langchain_merger}, to retrieve relevant scientific evidence from an academic knowledge base. The final output combines the verification results with corresponding evidence-based explanations.

\subsection{Data}

To construct the COVID-19 knowledge base, we created a contextual dataset by querying “COVID-19” and “SARS-CoV-2” in the PubMed and Scopus databases. The dataset includes metadata—titles, abstracts, author names, and keywords—from 126,984 peer-reviewed papers published between January 1, 2020, and January 1, 2024. Notably, 31\% of the papers were published in 2021 or 2022, and 22\% in 2023, ensuring up-to-date and well-established insights. Our previous study confirmed the effectiveness of this dataset in supporting reliable COVID-19 fact-checking.

To evaluate the SAFE system, we collected 50 fake COVID-19 news articles from the Science Feedback platform \cite{sciencefeedback}, a science-focused fact-checking site, and converted them into PDF format to form the evaluation dataset. The articles, ranging from 2 to 17 pages in length, covered a variety of COVID-19 topics, including vaccines, the origins of the virus, treatments, and more. Table \ref{tab:covid_fake_news} presents representative example titles of the fake news articles.

\begin{table*}[ht!]
\centering 
\caption{Example titles of the COVID-19 fake news articles}
\label{tab:covid_fake_news} 
\begin{tabular}{l p{0.7\textwidth}}
\toprule
\textbf{Category} & \textbf{Example titles of the COVID-19 fake news articles} \\
\midrule
Alternative Remedies & Shanghai Government Officially Recommends Vitamin C for COVID-19 \\
\addlinespace 
Virus Origin Conspiracy & The Coronavirus May Have Leaked From a Lab \\
\addlinespace
Mask Misinformation & New Evidence Shows Wearing Face Mask Can Help Coronavirus Enter the Brain and Pose More Health Risk, Warn Expert \\
\addlinespace
Data Manipulation Conspiracy & Johns Hopkins Study Mysteriously Disappears after It Revealed, in Spite of COVID, No More Deaths in 2020 Than in Prior Years \\
\addlinespace
Vaccine Misinformation & ‘What I’ve Seen in the Last 2 Years Is Unprecedented’ Physician on COVID Vaccine Side Effects on Pregnant Women \\
\bottomrule
\end{tabular}
\end{table*}

\subsection{Data Preprocessing}

\begin{figure}[ht!]
    \centering
    \includegraphics[width=\linewidth]{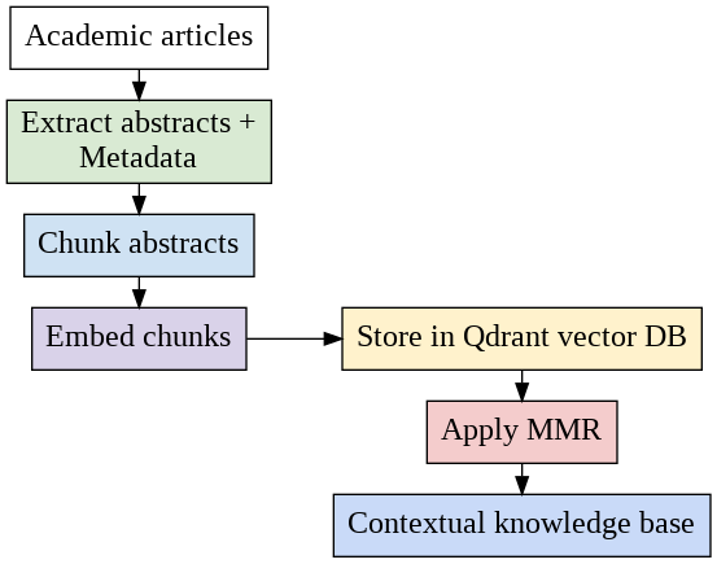}
    \caption{Figure 2. Data Preprocessing of RAG}
        \label{fig: Data Preprocessing}
\end{figure}

Figure \ref{fig: Data Preprocessing} illustrates the data preprocessing pipeline for building the contextual knowledge base used in the RAG system. For each academic article, we extracted the title, abstract, authors, and publication date. Abstracts were used as the primary content source due to their structured format, informativeness, and lower computational cost compared to full texts. Each abstract was segmented into overlapping text chunks, which, along with their metadata, were stored to provide contextual cues during retrieval. These chunks were embedded into high-dimensional vectors and indexed in a Qdrant-based vector store \cite{qdrant}. To enhance retrieval quality, we applied Maximal Marginal Relevance (MMR) with a similarity threshold of 0.8, balancing relevance and diversity among results.

For the 50 fake news articles in the evaluation dataset, we conducted manual fact-checking to extract and annotate COVID-19-related claims by public health professionals. The number of annotated claims per article had a mean of 4.922 and a standard deviation of 3.186. The annotation was guided by scientific knowledge retrieved from authoritative sources. These annotated claims were regarded as ground truth for evaluating the SAFE system, which aims to automatically identify all annotated COVID-19-related claims in the 50 articles, trying to classify them into five categories consistent with expert annotations—true (n = 45, 14.1\%), partly true (n = 46, 14.4\%), false (n = 86, 27.0\%), partly false (n = 7, 2.2\%), and misleading (n = 67, 21.0\%) (factually correct but contextually distorted)—and generate evidence-based explanations for each classification.

\subsection{Model Architecture}
\paragraph{Base Model}
We used OpenAI’s GPT-4o-mini as the foundational model for the RAG system, as it is the default model provided to users on the free tier of ChatGPT, offering a widely accessible and cost-effective AI solution. Within the Self-RAG (SRAG) framework, the GPT-4o model serves as both an automated evaluator and a response refinement module, enabling iterative assessment and enhancement of generated outputs.
\paragraph{Embedding Models}
Two distinct embedding models were applied in the SAFE systems. We combined the “text-embedding-3-small” model provided by OpenAI and the “NeuML/pubmedbert-base-embeddings” model from Hugging Face \cite{huggingface} to support the LOTR-RAG framework in the SAFE system while maintaining relatively quick response and low computational cost.
\paragraph{LOTR-RAG}
A naïve RAG system combines an LLM with a single vector-store retriever to ground responses in external knowledge, reducing hallucinations and improving contextual relevance. However, relying on one retriever may introduce bias in long-form tasks, as critical evidence can be missed and nuances overlooked.
To overcome this, we applied the LOTR-RAG system \cite{langchain_merger}. 

\begin{figure}[ht!]
    \centering
    \includegraphics[width=\linewidth]{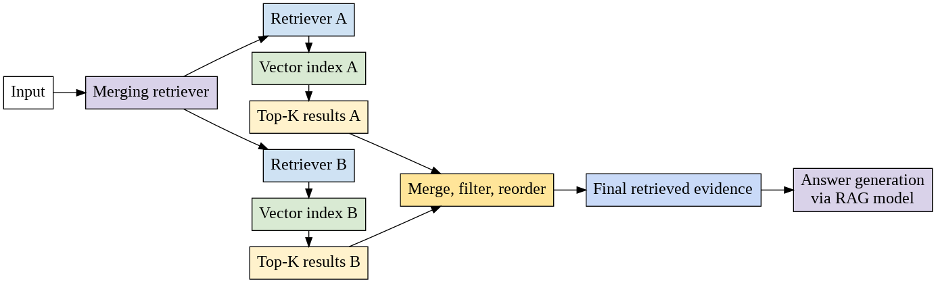}
    \caption{Figure 3. Workflow of LOTR-RAG}
        \label{fig: LOTR-RAG}
\end{figure}

As shown in Figure \ref{fig: LOTR-RAG}, LOTR-RAG employs a merging retriever that integrates two distinct embedding models, enabling filtration and reordering across two independently built vector indices from the same dataset. This design retrieves more comprehensive evidence, resulting in responses that are more accurate, coherent, and better aligned with the query than those generated by naïve RAG.

Despite enhancements, single-LLM systems, such as LOTR-RAG, may be limited by the absence of a self-reflective mechanism, potentially reducing their reliability in complex tasks, such as fact-checking \cite{li2025use}. To explore whether a self-reflective architecture could enhance performance, we introduced SRAG—a multi-agent framework that employs multiple LLMs to enable dynamic evaluation and iterative refinement.

\begin{figure}[ht!]
    \centering
    \includegraphics[width=\linewidth]{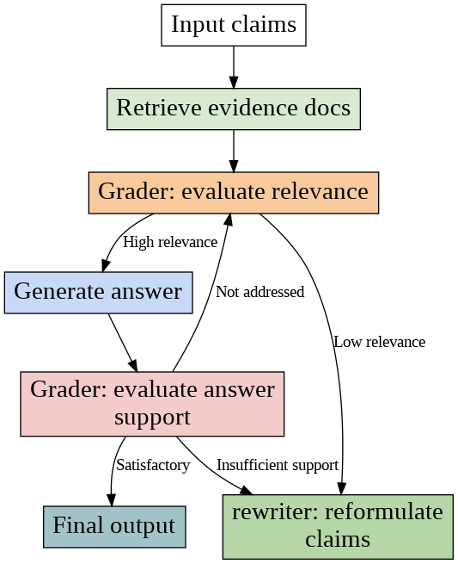}
    \caption{Figure 4. SRAG Framework}
        \label{fig: SRAG}
\end{figure}

\paragraph{Self-RAG}
Figure \ref{fig: SRAG} illustrates the SRAG framework \cite{asai2023self}, which uses GPT-4o as both a grader and rewriter, offering critical evaluation based on stronger general capabilities. After retrieving scientific evidence, the grader assesses its relevance to the input claims. If relevance is low, the rewriter reformulates the claims—preserving their meaning—to improve retrieval. This iterative cycle continues until high-relevance evidence is obtained, triggering answer generation.

Post-generation, another LLM grader evaluates whether the response sufficiently addresses the fact-checking task and is properly supported by evidence. If not, the system either rewrites the claim or regenerates the answer until a satisfactory result is produced. Through this self-reflective loop, SRAG can enhance the long-form COVID-19 fact-checking by ensuring internal validation before delivering outputs.

\subsection{Prompt Engineering}

Appendix \ref{appendix:prompt_engineering} presents the prompts utilized in this study. For claim extraction, the LLM is instructed to carefully identify and extract COVID-19-related claims from a single long-form article. For answer generation, the LLM performs fact-checking based on the retrieved documents from the contextual dataset. Within the agentic framework of SRAG, the LLM is further tasked with evaluating the relevance of retrieved evidence and the adequacy of generated responses.

\subsection{Evaluation}

To evaluate the performances of the SAFE systems in long-form COVID-19 fact-checking. We applied both objective and subjective evaluation methods. 

For objective evaluation, we used RAGAS \cite{ragas_docs}, an NLP evaluation toolkit, to measure the semantic similarity and factual consistency between SAFE system outputs and ground truths. Semantic similarity reflects the degree of alignment between a generated response and the reference answer, scored between 0 and 1. It is calculated by embedding both texts using the same model and computing their cosine similarity \cite{ragas_semantic}. Consistency assesses factual accuracy at the statement level by prompting an LLM to extract claims from both texts and applying natural language inference (NLI) to evaluate their overlap \cite{ragas_factual}. Higher scores on both metrics indicate stronger semantic and factual alignment. 

For subjective evaluation, we designed a 5-point Likert-scale questionnaire (0–4) to assess the usefulness, clearness, and authenticity of the fact-checking outputs. Usefulness was rated by the statement: “The LLM-generated answer is useful for me to do fact-checking with this article.” Clearness was assessed with: “The LLM-generated answer is clear to understand for me.” Authenticity was measured by: “The LLM-generated answer has sufficient evidence to be convincing for me to believe its explanation.” Four team members independently rated each output, and the average scores across dimensions were recorded as the final subjective evaluation

\subsection{Ethical Consideration}
This study utilized publicly available data containing no identifiable information. As such, it was exempt from institutional review board (IRB) review following federal regulations for the protection of human research subjects \cite{ohrp2025}.

\section{Results}

\begin{figure}[ht!]
    \centering
    \includegraphics[width=\linewidth]{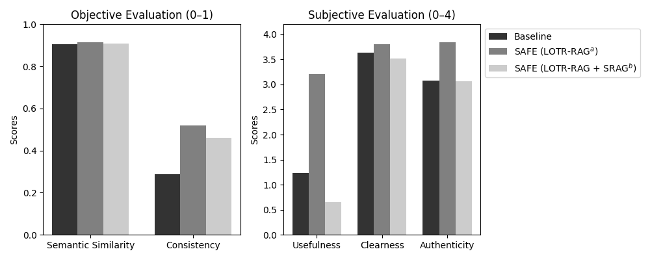}
    \caption{Figure 5. igure 5. Results of long-form COVID-19 fact-checking}
        \label{fig: Results}
\end{figure}

Figure \ref{fig: Results} shows the enhanced performance of our SAFE systems in long-form COVID-19 fact-checking. All improvements over the baseline were statistically significant (p < 0.001; see Multimedia Appendix \ref{appendix:pairwise_comparisons}). Objectively, all systems achieved similar semantic similarity scores, but SAFE (LOTR-RAG) attained the highest consistency (0.629), outperforming SAFE (LOTR-RAG+SRAG) (0.577) and the baseline (0.279). Subjectively, SAFE (LOTR-RAG) also led across all dimensions on the 0–4 Likert scale, with scores of 3.640 (usefulness), 3.800 (clearness), and 3.526 (authenticity), while the baseline scored lowest throughout. These findings indicate that both SAFE systems significantly improved LLM performance on the 50-article fake news dataset. However, adding the SRAG module did not enhance results further and showed slight declines across most metrics, except for clearness.

Appendix \ref{appendix:Examples of long-form COVID-19 fact-checking} presents an example of how SAFE systems enhance long-form COVID-19 fact-checking. Using a fake news article titled Cannabis is more effective at preventing and treating COVID-19 than hydroxychloroquine, we extracted and verified seven COVID-19-related claims as ground truths. The baseline GPT-4o-mini model failed to detect most of the relevant claims—this highlights the “lost in the middle” issue, where LLMs often miss key information in the middle of long inputs. In contrast, both SAFE systems successfully detected all ground-truth claims. The baseline model also produced hallucinations, including inaccurate facts and fabricated sources, while the SAFE systems substantially improved fact-checking quality. Among them, SAFE (LOTR-RAG) delivered the most accurate, evidence-based responses, underscoring its effectiveness for reliable long-form COVID-19 fact-checking. 

\section{Discussion}

This study evaluates the effectiveness of the multi-agent system (SAFE) in improving the accuracy and reliability of long-form COVID-19 fact-checking. Our approach tackles two key challenges in automated long-form fact-checking. First, large language models often generate inaccuracies and hallucinations. A notable example is the “lost in the middle” phenomenon, where critical information buried in the middle of lengthy text is overlooked. Second, traditional fact-checking—whether done manually or with conventional machine-learning methods—remains costly. By integrating a scientific dataset of approximately 130,000 peer-reviewed articles with GPT-4o-mini and implementing a multi-agent framework, we observed substantial improvement in both objective and subjective evaluation metrics. The SAFE improved factual consistency scores by 125\% over baseline while keeping (per-query) cost < \$0.05, demonstrating notable cost-efficiency.

In this study, the SAFE (LOTR-RAG + SRAG) system, which incorporates a self-refinement mechanism, did not yield further improvements over the SAFE (LOTR-RAG) system. While query rewriting through self-refinement may enhance retrieval, it can also compromise critical information necessary for accurate fact-checking of long-form COVID-19 articles.  For instance, the LLM rewriter may transform “5G is the origin of the virus” to “the human-made origin of the COVID-19 virus”, overlooking a key point of the original claim. Given that self-refinement and reflection are recognized as promising strategies to enhance the LLM systems \cite{renze2024self}, future work should explore more reliable methods to integrate self-reflective capabilities into long-form fact-checking tasks within multi-agent LLM frameworks.

During the COVID-19 pandemic, the widespread of misinformation and disinformation has severely undermined public health efforts by diminishing trust in science and fostering dangerous behaviors \cite{ferreira2022impact}. Furthermore, the sheer scale of fake news circulating online continues to outstrip the ability of manual approaches to manage its spread \cite{bateman2024countering}. Traditional automated approaches, such as building or fine-tuning deep learning models to classify long-text fake news \cite{choi2024fact, aggarwal2020classification}, have shown some effectiveness. However, conventional deep learning models often fall short when faced with the nuances of natural language processing and the detailed explanations required to debunk fake news \cite{oad2024fake}. Although LLM-based systems are considered a promising approach to fact-checking \cite{singhal2024evidence}, they struggle with long-form tasks and hallucinations, which ultimately lower the quality of long-form COVID-19 fact-checking.

\subsection{Theoretical Implication}

Integrating advanced RAG frameworks with LLM agents marks a significant advancement in automated fact-checking. Previous studies have explored the effectiveness of RAG-equipped LLMs in domains such as political claims \cite{khaliq2024ragar}. In our published work, we achieved over 97\% accuracy on short-form COVID-19 fact-checking tasks using RAG systems \cite{li2025use}. Unlike prior research that primarily focused on short, sentence-level claims, our work targets long-form news article fact-checking to address the challenges LLMs face in processing long-form tasks. In this study, with proper preprocessing of long documents, issues like the “lost in the middle” effect are largely mitigated, resulting in a workflow that remains both efficient and cost-effective. LOTR-RAG further enhances performance by combining diverse embedding models to retrieve more comprehensive contextual passages from external knowledge bases, thereby reducing information loss and mitigating hallucinations during complex tasks. The result is accurate, coherent answers tailored to domain-specific needs—health fact-checking in particular \cite{asai2023self}. Building on this foundation, our SAFE system targets the COVID-19 “infodemic.” It chunks lengthy articles, uses a specialized LLM agent to extract key claims, and applies LOTR-RAG techniques to verify them. The outcome is a state-of-the-art, easily accessible solution for long-form fact-checking that meets public health demands for rigorous, evidence-based explanations.

\subsection{Future Work}

AI-driven health fact-checking is widely regarded as essential for combating infodemics, but it still struggles with long-form content \cite{sahitaj2025towards, kim2024can}. Designed for such texts, SAFE delivers accurate COVID-19 fact-checking and holds potential for extension to other domains like cancer, HIV/AIDS and nutrition. Besides, future work should address the complexity of multimodal content, nuanced context, and real-world user engagement. Emerging advances—such as reinforcement-based fine-tuning and graph-augmented retrieval—offer promising directions. To realize this potential will require collaboration between public health experts and AI developers to ensure ethical deployment, data protection, and alignment with public health objectives.

\subsection{Limitations}

Despite its strengths, the SAFE system has limitations. First, the LLM agent struggles with non-textual elements like images and graphs, which may contain critical information, and can misinterpret articles with complex visual or structural layouts. Second, the external academic knowledge base is incomplete due to language barriers, regional disparities, and the lack of unpublished or timely clinical data. Although LOTR-RAG improves retrieval accuracy, it remains constrained by the quality and scope of the dataset. Third, since the context dataset is built from academic abstracts, it may not capture the vernacular or stylistic nuances of social media, limiting the system’s accuracy in such domains. These challenges may affect responsiveness in real-time applications. Addressing them is essential to improve the quality of automated long-form fact-checking in public health settings.

\section{Conclusion}

This study demonstrates that the SAFE system—an integration of advanced RAG frameworks with a multi-agent architecture—significantly enhances the accuracy and explanatory quality of automated long-form fact-checking. By addressing key limitations of LLMs, including difficulties with long-form content, factual inaccuracies, and hallucinations, while also reducing the resource intensiveness of traditional fact-checking methods, the SAFE system improves the reliability of long-form COVID-19 fact-checking outputs. These advancements position RAG- and LLM-based approaches as scalable and effective solutions for combating misinformation, particularly in the context of public health emergencies.

\bibliography{custom}

\begin{thebibliography}{43}
\providecommand{\natexlab}[1]{#1}

\bibitem[{Aggarwal et~al.(2020)Aggarwal, Chauhan, Kumar, Mittal, and Verma}]{aggarwal2020classification}
Archita Aggarwal, Abhishek Chauhan, D~Kumar, M~Mittal, and S~Verma. 2020.
\newblock \href {https://doi.org/10.4108/eai.13-7-2018.163973} {Classification of fake news by fine-tuning deep bidirectional transformers-based language model}.
\newblock In \emph{Proceedings of SIS}. EAI.

\bibitem[{Andrew(2024)}]{andrew2024understanding}
Amos Andrew. 2024.
\newblock \href {https://doi.org/10.4082/kjfm.23.0274} {Understanding the "infodemic" threat: A case study of the {COVID-19} pandemic}.
\newblock \emph{Korean Journal of Family Medicine}, 45(4):183--188.

\bibitem[{Asai et~al.(2023)Asai, Wu, Wang, Sil, and Hajishirzi}]{asai2023self}
Akari Asai, Zeqiu Wu, Yizhong Wang, Avirup Sil, and Hannaneh Hajishirzi. 2023.
\newblock {Self-RAG}: Learning to retrieve, generate, and critique through self-reflection.
\newblock \emph{arXiv preprint arXiv:2310.11511}.

\bibitem[{Atehortua and Patino(2021)}]{atehortua2021covid}
Natalia~A Atehortua and Sandra Patino. 2021.
\newblock \href {https://doi.org/10.1093/heapro/daaa140} {{COVID-19}, a tale of two pandemics: novel coronavirus and fake news messaging}.
\newblock \emph{Health Promotion International}, 36(2):524--534.

\bibitem[{Augenstein et~al.(2024)Augenstein, Baldwin, Cha et~al.}]{augenstein2024factuality}
Isabelle Augenstein, Timothy Baldwin, Meeyoung Cha, and 1 others. 2024.
\newblock \href {https://doi.org/10.1038/s42256-024-00881-z} {Factuality challenges in the era of large language models and opportunities for fact-checking}.
\newblock \emph{Nature Machine Intelligence}, 6:852--863.

\bibitem[{Baker et~al.(2024)Baker, Raut, Shaier, and Hunter}]{baker2024lost}
Gabriel~A Baker, Ansh Raut, Sagnik Shaier, and Lawrence~E Hunter. 2024.
\newblock Lost in the middle, and in-between: Enhancing language models' ability to reason over long contexts in multi-hop {QA}.
\newblock \emph{arXiv preprint arXiv:2412.10079}.

\bibitem[{Bateman and Jackson(2024)}]{bateman2024countering}
Jon Bateman and Dean Jackson. 2024.
\newblock Countering disinformation effectively: An evidence-based policy guide.
\newblock Technical report, Carnegie Endowment for International Peace.

\bibitem[{Borges~do Nascimento et~al.(2022)Borges~do Nascimento, Pizarro, Almeida et~al.}]{borges2022infodemics}
Ivana~Jesus Borges~do Nascimento, Andre~B Pizarro, J{\'e}ssica~M Almeida, and 1 others. 2022.
\newblock \href {https://doi.org/10.2471/BLT.21.287654} {Infodemics and health misinformation: a systematic review of reviews}.
\newblock \emph{Bulletin of the World Health Organization}, 100(9):544--561.

\bibitem[{Cai et~al.(2024)Cai, Tan, Song, Sun, Jiang, Xu, Zhang, and Gu}]{cai2024forag}
Tianshu Cai, Zhe Tan, Xiaolan Song, Tianshi Sun, Jintang Jiang, Yikun Xu, Yutao Zhang, and Jiquan Gu. 2024.
\newblock \href {https://doi.org/10.1145/3637528.3672065} {{FoRAG}: Factuality-optimized retrieval augmented generation for web-enhanced long-form question answering}.
\newblock In \emph{Proceedings of the 30th ACM SIGKDD Conference on Knowledge Discovery and Data Mining}, pages 199--210.

\bibitem[{Choi and Ferrara(2024)}]{choi2024fact}
Eun~Cheol Choi and Emilio Ferrara. 2024.
\newblock \href {https://doi.org/10.1145/3589335.3651504} {{FACT-GPT}: Fact-checking augmentation via claim matching with {LLMs}}.
\newblock In \emph{Companion Proceedings of the ACM Web Conference 2024}, pages 883--886.

\bibitem[{Ferreira~Caceres et~al.(2022)Ferreira~Caceres, Sosa, Lawrence et~al.}]{ferreira2022impact}
Maria~M Ferreira~Caceres, Jhon~P Sosa, Jhon~A Lawrence, and 1 others. 2022.
\newblock \href {https://doi.org/10.3934/publichealth.2022018} {The impact of misinformation on the {COVID-19} pandemic}.
\newblock \emph{AIMS Public Health}, 9(2):262--277.

\bibitem[{Gao et~al.(2023)Gao, Xiong, Gao, Jia, Pan, Bi, Dai, Sun, Wang, and Wang}]{gao2023retrieval}
Yunfan Gao, Yun Xiong, Xinyu Gao, Kangxiang Jia, Jinliu Pan, Yuxi Bi, Yi~Dai, Jiawei Sun, Meng Wang, and Haofen Wang. 2023.
\newblock Retrieval-augmented generation for large language models: A survey.
\newblock \emph{arXiv preprint arXiv:2312.10997}.

\bibitem[{{Hugging Face}(2025)}]{huggingface}
{Hugging Face}. 2025.
\newblock {Hugging Face}.
\newblock \url{https://huggingface.co/}.

\bibitem[{Ji et~al.(2023)Ji, Lee, Frieske, Yu, Su, Xu, Ishii, Bang, Madotto, and Fung}]{ji2023survey}
Ziwei Ji, Nayeon Lee, Rita Frieske, Tiezheng Yu, Dan Su, Yan Xu, Etsuko Ishii, Yejin~J Bang, Andrea Madotto, and Pascale Fung. 2023.
\newblock \href {https://doi.org/10.1145/3571730} {Survey of hallucination in natural language generation}.
\newblock \emph{ACM Computing Surveys}, 55(12):1--38.

\bibitem[{Khaliq et~al.(2024)Khaliq, Chang, Ma, Pflugfelder, and Mileti{\'c}}]{khaliq2024ragar}
Muhammad~Abdullah Khaliq, Ping Chang, Minchen Ma, Bernhard Pflugfelder, and Filip Mileti{\'c}. 2024.
\newblock {RAGAR}, your falsehood {RADAR}: {RAG-augmented} reasoning for political fact-checking using multimodal large language models.
\newblock \emph{arXiv preprint arXiv:2404.12065}.

\bibitem[{Kim et~al.(2024)Kim, Lee, Huang, Chan, Li, and Ji}]{kim2024can}
Kunjal Kim, Saemi Lee, Kaize Huang, Hoi~Ping Chan, Meng Li, and Heng Ji. 2024.
\newblock Can {LLMs} produce faithful explanations for fact-checking? towards faithful explainable fact-checking via multi-agent debate.
\newblock \emph{arXiv preprint arXiv:2402.07401}.

\bibitem[{Kolluri et~al.(2022)Kolluri, Liu, and Murthy}]{kolluri2022covid}
Naitik Kolluri, Yuci Liu, and Dhiraj Murthy. 2022.
\newblock \href {https://doi.org/10.2196/38756} {{COVID-19} misinformation detection: Machine-learned solutions to the infodemic}.
\newblock \emph{JMIR Infodemiology}, 2(2):e38756.

\bibitem[{{LangChain}(2025)}]{langchain_merger}
{LangChain}. 2025.
\newblock {Merger Retriever}.
\newblock \url{https://python.langchain.com/docs/integrations/retrievers/merger_retriever/}.

\bibitem[{Leite et~al.(2025)Leite, Razuvayevskaya, Scarton, and Bontcheva}]{leite2025cross}
Jo{\~a}o~A Leite, Olga Razuvayevskaya, Carolina Scarton, and Kalina Bontcheva. 2025.
\newblock \href {https://doi.org/10.1145/3701716.3715535} {A cross-domain study of the use of persuasion techniques in online disinformation}.
\newblock In \emph{Companion Proceedings of the ACM on Web Conference 2025}, pages 1100--1103. ACM.

\bibitem[{Lewis et~al.(2020)Lewis, Perez, Piktus, Petroni, Karpukhin, Goyal, K{\"u}ttler, Lewis, Yih, Rockt{\"a}schel, Riedel, and Kiela}]{lewis2020retrieval}
Patrick Lewis, Ethan Perez, Aleksandra Piktus, Fabio Petroni, Vladimir Karpukhin, Naman Goyal, Heinrich K{\"u}ttler, Mike Lewis, Wen-tau Yih, Tim Rockt{\"a}schel, Sebastian Riedel, and Douwe Kiela. 2020.
\newblock Retrieval-augmented generation for knowledge-intensive {NLP} tasks.
\newblock \emph{arXiv preprint arXiv:2005.11401}.

\bibitem[{Li et~al.(2025)Li, Huang, Ji, Yang, and An}]{li2025use}
Haotian Li, Jingcheng Huang, Meichen Ji, Yixuan Yang, and Ru~An. 2025.
\newblock \href {https://doi.org/10.2196/66098} {Use of retrieval-augmented large language model for {COVID-19} fact-checking: Development and usability study}.
\newblock \emph{Journal of Medical Internet Research}, 27:e66098.

\bibitem[{Momii et~al.(2024)Momii, Takiguchi, and Ariki}]{momii2024rag}
Yuto Momii, Tetsuya Takiguchi, and Yasuo Ariki. 2024.
\newblock {RAG-Fusion} based information retrieval for fact-checking.
\newblock In \emph{Proceedings of the Seventh Fact Extraction and VERification Workshop (FEVER)}, pages 47--54. Association for Computational Linguistics.

\bibitem[{Oad et~al.(2024)Oad, Farooq, Zafar, Akram, Zhou, and Dong}]{oad2024fake}
Arjun Oad, Muhammad~Hassan Farooq, Adeel Zafar, Beenish~A Akram, Rongting Zhou, and Feng Dong. 2024.
\newblock \href {https://doi.org/10.1109/ACCESS.2024.3491376} {Fake news classification methodology with enhanced {BERT}}.
\newblock \emph{IEEE Access}, 12:164491--164502.

\bibitem[{{Office for Human Research Protections}(2025)}]{ohrp2025}
{Office for Human Research Protections}. 2025.
\newblock {45 CFR 46}.
\newblock \url{https://www.hhs.gov/ohrp/regulations-and-policy/regulations/45-cfr-46/index.html}.

\bibitem[{{Qdrant}(2025)}]{qdrant}
{Qdrant}. 2025.
\newblock {Qdrant: Vector similarity search engine}.
\newblock \url{https://qdrant.tech/}.

\bibitem[{{Ragas}(2025{\natexlab{a}})}]{ragas_docs}
{Ragas}. 2025{\natexlab{a}}.
\newblock Documentation.
\newblock \url{https://docs.ragas.io/en/stable/}.

\bibitem[{{Ragas}(2025{\natexlab{b}})}]{ragas_factual}
{Ragas}. 2025{\natexlab{b}}.
\newblock Factual correctness metric.
\newblock \url{https://docs.ragas.io/en/stable/concepts/metrics/available_metrics/factual_correctness/}.

\bibitem[{{Ragas}(2025{\natexlab{c}})}]{ragas_semantic}
{Ragas}. 2025{\natexlab{c}}.
\newblock Semantic similarity metric.
\newblock \url{https://docs.ragas.io/en/stable/concepts/metrics/available_metrics/semantic_similarity/}.

\bibitem[{Renze and Guven(2024)}]{renze2024self}
Maximilian Renze and Enes Guven. 2024.
\newblock \href {https://doi.org/10.1109/FLLM63129.2024.10852493} {Self-reflection in {LLM} agents: Effects on problem-solving performance}.
\newblock In \emph{2024 IEEE International Conference on Foundation and Large Language Models (FLLM)}. IEEE.

\bibitem[{{Reuters}(2024)}]{reuters2024misleading}
{Reuters}. 2024.
\newblock Misleading data used to claim {COVID} vaccines do more harm than good.
\newblock \url{https://www.reuters.com/fact-check/misleading-data-used-claim-covid-vaccines-do-more-harm-than-good-2024-03-21/}.
\newblock Accessed: 2025-06-10.

\bibitem[{Rodrigues et~al.(2024)Rodrigues, Newell, Babu, Chatterjee, Sandhu, and Gupta}]{rodrigues2024social}
Filipe Rodrigues, Ryan Newell, Giridhara~R Babu, Tanusree Chatterjee, Nav~K Sandhu, and Lovesh Gupta. 2024.
\newblock \href {https://doi.org/10.1016/j.hlpt.2024.100846} {The social media infodemic of health-related misinformation and technical solutions}.
\newblock \emph{Health Policy and Technology}, 13(2):100846.

\bibitem[{Russo et~al.(2024)Russo, Menini, Staiano, and Guerini}]{russo2024face}
Daniele Russo, Stefano Menini, Jacopo Staiano, and Marco Guerini. 2024.
\newblock Face the facts! evaluating {RAG-based} fact-checking pipelines in realistic settings.
\newblock \emph{arXiv preprint arXiv:2412.15189}.

\bibitem[{Sahitaj et~al.(2025)Sahitaj, Maab, Yamagishi, Kolanowski, M{\"o}ller, and Schmitt}]{sahitaj2025towards}
Pjeter Sahitaj, Ines Maab, Junichi Yamagishi, Jurek Kolanowski, Sebastian M{\"o}ller, and Vincent Schmitt. 2025.
\newblock Towards automated fact-checking of real-world claims: Exploring task formulation and assessment with {LLMs}.
\newblock \emph{arXiv preprint}.

\bibitem[{{Science Feedback}(2025)}]{sciencefeedback}
{Science Feedback}. 2025.
\newblock {Science Feedback}.
\newblock \url{https://science.feedback.org/}.

\bibitem[{Sevgili et~al.(2024)Sevgili, Nikishina, Yimam, Semmann, and Biemann}]{sevgili2024uhh}
{\"O}zge Sevgili, Irina Nikishina, Seid~M Yimam, Merit Semmann, and Chris Biemann. 2024.
\newblock {UHH} at {AVeriTeC}: {RAG} for fact-checking with real-world claims.
\newblock In \emph{Proceedings of the Seventh Fact Extraction and VERification Workshop (FEVER)}, pages 55--63. Association for Computational Linguistics.

\bibitem[{Singhal et~al.(2024)Singhal, Patwa, Patwa, Chadha, and Das}]{singhal2024evidence}
Raj Singhal, Parth Patwa, Pavan Patwa, Aman Chadha, and Amitava Das. 2024.
\newblock Evidence-backed fact checking using {RAG} and few-shot in-context learning with {LLMs}.
\newblock \emph{arXiv preprint arXiv:2408.12060}.

\bibitem[{Tang et~al.(2024)Tang, Laban, and Durrett}]{tang2024minicheck}
Lifan Tang, Philipp Laban, and Greg Durrett. 2024.
\newblock \href {https://doi.org/10.48550/arXiv.2404.10774} {{MiniCheck}: Efficient fact-checking of {LLMs} on grounding documents}.
\newblock \emph{arXiv preprint arXiv:2404.10774}.

\bibitem[{{The Lancet Infectious Diseases}(2020)}]{thelancet2020infodemic}
{The Lancet Infectious Diseases}. 2020.
\newblock \href {https://doi.org/10.1016/S1473-3099(20)30565-X} {The covid-19 infodemic}.
\newblock \emph{The Lancet Infectious Diseases}, 20(8):875.

\bibitem[{Wang et~al.(2022)Wang, Bye, Bales et~al.}]{wang2022understanding}
Yuxi Wang, Jessica Bye, Kate Bales, and 1 others. 2022.
\newblock \href {https://doi.org/10.1136/bmj-2022-070331} {Understanding and neutralising {COVID-19} misinformation and disinformation}.
\newblock \emph{BMJ}, 379:e070331.

\bibitem[{Wei et~al.(2024)Wei, Yang, Song, Lu, Hu, Huang, Tran, Peng, Liu, Huang, Du, and Le}]{wei2024long}
Jerry Wei, Chen Yang, Xuezhi Song, Yifan Lu, Nathan Hu, Jialu Huang, Dai Tran, Da~Peng, Ruofei Liu, Denny Huang, Charles Du, and Quoc~V Le. 2024.
\newblock Long-form factuality in large language models.
\newblock \emph{arXiv preprint arXiv:2403.18802}.

\bibitem[{Wilhelm et~al.(2023)Wilhelm, Ballalai, Belanger et~al.}]{wilhelm2023measuring}
Elisabeth Wilhelm, Isabella Ballalai, M.~E. Belanger, and 1 others. 2023.
\newblock \href {https://doi.org/10.2196/44207} {Measuring the burden of infodemics: summary of the methods and results of the fifth {WHO} infodemic management conference}.
\newblock \emph{JMIR Infodemiology}, 3:e44207.

\bibitem[{Yang et~al.(2021)Yang, Shin, Zhou, Huang-Isherwood, Lee, Dong, Kim, Zhang, Sun, Li, Nan, Zhen, and Liu}]{yang2021battleground}
Ai~Yang, Jiyoun Shin, Angela Zhou, Katherine~M Huang-Isherwood, Emily Lee, Catherine Dong, H.~M. Kim, Yimeng Zhang, Jialing Sun, Yifei Li, Yixuan Nan, Lin Zhen, and Wei Liu. 2021.
\newblock \href {https://doi.org/10.37016/mr-2020-78} {The battleground of {COVID-19} vaccine misinformation on {Facebook}: Fact checkers vs. misinformation spreaders}.
\newblock \emph{Harvard Kennedy School (HKS) Misinformation Review}.

\bibitem[{Yang et~al.(2024)Yang, Peng, Wang, Wen, and Zhang}]{yang2024llm}
Yi~Yang, Qirui Peng, Jing Wang, Yilong Wen, and Wei Zhang. 2024.
\newblock {LLM-based} multi-agent systems: Techniques and business perspectives.
\newblock \emph{arXiv preprint arXiv:2411.14033}.

\end{thebibliography}


\appendix
\clearpage
\section{Prompt Engineering}
\label{appendix:prompt_engineering}

\begin{table*}[ht!]
\centering
\caption{Prompts used in experiments.}
\begin{tabular}{l p{0.7\textwidth}}
\toprule
\textbf{Objective} & \textbf{Prompt} \\
\midrule

Generate answers & 
You are an expert for the COVID-19 fact-checking tasks. Based on pieces of retrieved context to detect if the claim is true or false. You will have to give me the title and author of the context you referred to in one sentence. If you don’t know the answer, just say that you don’t know. Keep the answer concise. \newline \newline
Claim: \texttt{\{question\}} \newline
Context: \texttt{\{context\}} \newline
Answer: \\

\addlinespace

Grade documents & 
You are a grader assessing the relevance of a retrieved document to a user question. If the document contains keywords related to the user question, grade it as relevant. It does not need to be a stringent test. The goal is to filter out erroneous retrievals. Give a binary score of “yes” or “no” score to indicate whether the document is relevant to the question. Provide the binary score as a JSON object with a single key, “score,” and no preamble or explanation. \newline \newline
Here is the retrieved document: \texttt{\{document\}} \newline
Here is the user question: \texttt{\{question\}} \\

\addlinespace

Grade answers & 
You are a grader assessing whether an answer is useful to resolve a question. Give a binary score of “yes” or “no” to indicate whether the answer is useful to resolve a question. Provide the binary score as a JSON object with a single key, “score,” and no preamble or explanation. \newline \newline
Here is the answer: \texttt{\{generation\}} \newline
Here is the question: \texttt{\{question\}} \\

\addlinespace

Rewrite claims &
You are a claim rewriter who converts an input claim to a better version that is optimized for vector store retrieval and fact-checking. Look at the input and try to reason about the underlying semantic intent meaning. \\

\bottomrule
\end{tabular}
\end{table*}


\clearpage
\section{Pairwise Comparisons}
\label{appendix:pairwise_comparisons}

\begin{table*}[htbp]
\centering
\caption{Pairwise Comparisons of Fact-Checking Quality Across Models.}
\adjustbox{max width=\textwidth}{%
\begin{tabular}{llcccccc}
\toprule
\textbf{Comparison} & \textbf{Metric} & \textbf{Model A Score} & \textbf{Model B Score} & \textbf{$\Delta$ (A $-$ B)} & \textbf{$p(t)$} & \textbf{$p(W)$} & \textbf{Significance} \\
\midrule

\multirow{2}{*}{Baseline vs. SAFE (LOTR-RAG)} 
& Semantic Similarity & 0.904 & 0.918 & +0.014 & 0.0179 & 0.0003 & * \\
& Consistency         & 0.287 & 0.521 & +0.233 & 0.0000 & 0.0000 & *** \\
\addlinespace

\multirow{2}{*}{SAFE (LOTR-RAG) vs. +SRAG} 
& Semantic Similarity & 0.918 & 0.916 & $-$0.002 & 0.1532 & 0.3622 & n.s. \\
& Consistency         & 0.521 & 0.460 & $-$0.061 & 0.0005 & 0.0012 & *** \\
\addlinespace

\multirow{2}{*}{Baseline vs. SAFE (LOTR-RAG + SRAG)} 
& Semantic Similarity & 0.904 & 0.916 & +0.012 & 0.0405 & 0.0026 & * \\
& Consistency         & 0.287 & 0.460 & +0.172 & 0.0000 & 0.0000 & *** \\

\bottomrule
\end{tabular}
}

\vspace{1em}
\begin{flushleft}
\small
\textbf{Note:} \\
$p(t)$: $p$-value from paired $t$-test \\
$p(W)$: $p$-value from Wilcoxon signed-rank test \\
Significance codes: *** $p < .001$, ** $p < .01$, * $p < .05$, n.s. = not significant \\
SAFE: System for Accurate Fact Extraction and Evaluation \\
RAG: Retrieval-Augmented Generation \\
LOTR-RAG: Lord of the Retrievers \\
SRAG: Self-RAG
\end{flushleft}
\end{table*}


\onecolumn 

\section{Multimedia Appendix 3. Examples of long-form COVID-19 fact-checking.}
\label{appendix:longform_covid_factcheck}

\begin{singlespace}
\begin{longtable}{>{\raggedright\arraybackslash}p{0.15\textwidth} >{\raggedright\arraybackslash}p{0.4\textwidth} >{\raggedright\arraybackslash}p{0.4\textwidth}}

\caption{Examples of long-form COVID-19 fact-checking.}
\label{tab:covid_factcheck} \\

\toprule
\textbf{Fact-checking} & \textbf{Extracted claims} & \textbf{Response of fact-checking} \\
\midrule
\endfirsthead

\caption[]{-- continued from previous page} \\
\toprule
\textbf{Fact-checking} & \textbf{Extracted claims} & \textbf{Response of fact-checking} \\
\midrule
\endhead

\midrule
\multicolumn{3}{r}{\small\textit{Continued on next page}} \\
\endfoot

\bottomrule
\endlastfoot

\multicolumn{3}{p{0.95\textwidth}}{\textbf{Title of article:} \textit{Cannabis is more effective at preventing and treating COVID-19 than hydroxychloroquine}} \\
\addlinespace[3pt]
\midrule

\multirow{7}{*}{\textbf{Ground truth}} 
& 1. Canadian scientists found 13 strains of Cannabis sativa that may help prevent or treat COVID-19.
& 1. False – There is no scientific evidence to support the claim that cannabis is more effective than hydroxychloroquine in preventing or treating COVID-19. \\
\cmidrule(lr){2-3}
& 2. Studies show hydroxychloroquine, promoted by the U.S. President, increases risks like death and heart issues in COVID-19 patients.
& 2. Partly true – While some studies have explored the potential antiviral effects of certain compounds found in cannabis, such as cannabinoids, there is currently no conclusive evidence that any strains of cannabis sativa can aid in the prevention or treatment of COVID-19. \\
\cmidrule(lr){2-3}
& 3. Cannabis is claimed to be more effective against COVID-19 than hydroxychloroquine.
& 3. Partly true – Hydroxychloroquine, originally developed to treat malaria and autoimmune conditions, has been associated with serious side effects, particularly when used outside of clinical guidelines. \\
\cmidrule(lr){2-3}
& 4. High-CBD cannabis extracts may affect ACE2 expression in COVID-19 target tissues.
& 4. False – Some in vitro studies suggest that certain high-CBD cannabis extracts may modulate ACE2 expression in epithelial tissues, but this is not equivalent to proven therapeutic effect, and results have not been replicated in animal or human studies. \\
\cmidrule(lr){2-3}
& 5. The cannabis extracts may hinder SARS-CoV-2 from entering human cells.
& 5. Partly true – The hypothesis is biologically plausible (reducing ACE2 might affect viral entry), but there is no direct evidence that cannabis extracts effectively block SARS-CoV-2 from entering human. \\
\cmidrule(lr){2-3}
& 6. A Harvard-led study linked hydroxychloroquine to higher death rates and severe heart problems.
& 6. False –The claim misrepresents the findings of studies regarding hydroxychloroquine, which suggest it may be associated with lower mortality and have low incidence of severe QTc prolongation. \\
\cmidrule(lr){2-3}
& 7. Donald Trump said he was taking hydroxychloroquine to prevent COVID-19, despite no peer-reviewed support for its safety or effectiveness.
& 7. True – Donald Trump publicly stated he was taking hydroxychloroquine as a preventive measure against COVID-19. At the time, there was no conclusive peer-reviewed evidence supporting its safety or efficacy for this purpose. \\
\midrule

\multirow{3}{*}{\textbf{Baseline}}
& 1. A team of scientists from Canada have identified 13 cannabis strains that can help prevent and treat COVID-19.
& 1. Misleading. While the research does suggest that certain cannabis extracts might interact with the ACE2 receptor, there is no definitive evidence to show that cannabis can prevent or treat COVID-19 in humans. Source: Cannabis research is still in very early stages. \\
\cmidrule(lr){2-3}
& 2. Hydroxychloroquine is dangerous and ineffective for treating COVID-19, with studies showing it increases mortality risk.
& 2. Partly true. Several studies indicated that hydroxychloroquine can increase risks of death and heart arrhythmias. Source: The article cites studies such as the one from Harvard Medical School, which analyzed over 96,000 patients. \\
\cmidrule(lr){2-3}
& 3. Cannabis has medical benefits for treating diseases ranging from viral infections to cancer.
& 3. Partly true. Cannabis and its compounds have shown some promise in managing symptoms of various conditions. Source: The FDA has approved cannabis-derived drugs for specific conditions like epilepsy (Epidiolex), but there is no conclusive evidence supporting cannabis as a treatment for COVID-19 or cancer itself. \\
\midrule

\multirow{8}{*}{\textbf{SAFE (LOTR-RAG)}}
& 1. A team of Canadian scientists identified at least 13 strains of Cannabis sativa that may aid in the prevention and treatment of COVID-19.
& 1. False. No specific study has conclusively identified 13 strains of Cannabis sativa as effective in preventing or treating COVID-19. Source: J. McGrail; L. Martín-Banderas; M. Durán-Lobato, 2022. \\
\cmidrule(lr){2-3}
& 2. Hydroxychloroquine, a drug touted by the U.S. President, has been shown in studies to have dangerous side effects.
& 2. Partly True. While hydroxychloroquine has been linked to increased mortality in certain studies, other studies reported it to be well-tolerated. Source: C. Axfors et al. (2021). \\
\cmidrule(lr){2-3}
& 3. Cannabis is claimed to be more effective at preventing and treating COVID-19 than hydroxychloroquine.
& 3. Partly true. While cannabis showed some benefits for hospitalized patients, asserting it is more effective than hydroxychloroquine misrepresents the available evidence. Source: J. Tan et al., 2021; C. M. Shover et al., 2022. \\
\cmidrule(lr){2-3}
& 4. High-CBD cannabis extracts may modulate ACE2 expression in tissues that are targets for COVID-19.
& 4. False. There is currently insufficient unbiased clinical evidence to support that CBD can effectively modulate ACE2 expression. Source: J. H. Khalsa; G. Bunt; S. B. Maggirwar; S. Kottilil, 2021. \\
\cmidrule(lr){2-3}
& 5. The identified cannabis extracts may make it more difficult for the SARS-CoV-2 coronavirus to enter host cells.
& 5. Partly true. There is currently no conclusive evidence directly demonstrating that specific cannabis extracts can effectively prevent SARS-CoV-2 from entering host cells. Source: J. McGrail; L. Martín-Banderas; M. Durán-Lobato, 2022. \\
\cmidrule(lr){2-3}
& 6. A study led by Harvard Medical School indicated that patients treated with hydroxychloroquine had an increased risk of mortality.
& 6. False. No study finds that hydroxychloroquine was associated with increased risk of mortality; rather, it was associated with lower mortality when adjusted for other variables. Source: L. Ayerbe; C. Risco-Risco; S. Ayis, 2020. \\
\cmidrule(lr){2-3}
& 7. Donald Trump claimed to be taking hydroxychloroquine as a preventative measure against COVID-19.
& 7. True. Donald Trump did claim to be taking hydroxychloroquine; however, peer-reviewed studies have indicated limited effectiveness. Source: D. Morgado-Carrasco; J. Ibaceta-Ayala; J. Piquero-Casals, 2022. \\
\cmidrule(lr){2-3}
& 8. Cannabis has not been associated with deaths and has potential medical benefits for various diseases.
& 8. Partly true. The available clinical evidence does not support its potential medical benefits for viral infections, including COVID-19. Source: J. H. Khalsa; S. B. Maggirwar; G. Bunt, 2021. \\
\midrule

\multirow{8}{*}{\textbf{SAFE (LOTR-RAG + SRAG)}}
& 1. A team of Canadian scientists identified at least 13 strains of Cannabis sativa.
& 1. Misleading. There is no specific confirmation of 13 unique strains of Cannabis sativa being identified for this purpose. Source: J. McGrail; L. Martín-Banderas; M. Durán-Lobato, 2022. \\
\cmidrule(lr){2-3}
& 2. Hydroxychloroquine has been shown in studies to have dangerous side effects, including increased mortality risks.
& 2. True. Hydroxychloroquine has been associated with significant side effects and is linked to increased mortality risk in patients with COVID-19. Source: C. Axfors et al., 2021. \\
\cmidrule(lr){2-3}
& 3. Cannabis is claimed to be more effective at preventing and treating COVID-19 than hydroxychloroquine.
& 3. False. There is no scientific evidence supporting the claim that cannabis is more effective than hydroxychloroquine in preventing or treating COVID-19. Source: D. S. K. Pathak; et al., 2020. \\
\cmidrule(lr){2-3}
& 4. High-CBD cannabis extracts may modulate ACE2 expression in tissues that are targets for COVID-19.
& 4. True. The claim is supported by research that explores the effects of high-CBD cannabis extracts on ACE2 expression. Source: R. R. J. van Kimmenade; et al., 2021. \\
\cmidrule(lr){2-3}
& 5. The identified cannabis extracts may make it more difficult for the SARS-CoV-2 coronavirus to enter host cells.
& 5. True. Identified cannabis extracts have shown potential antiviral properties that may inhibit the SARS-CoV-2 coronavirus's ability to penetrate host cells. Source: N. K. Jha et al., 2021. \\
\cmidrule(lr){2-3}
& 6. A study led by Harvard Medical School indicated that patients treated with hydroxychloroquine had an increased risk of mortality.
& 6. True. A study examining hydroxychloroquine found that patients who took the drug had an elevated risk of mortality. Source: M. Luo; B. Wu; Y. Li; F. Wu, 2022. \\
\cmidrule(lr){2-3}
& 7. Donald Trump claimed to be taking hydroxychloroquine as a preventative measure.
& 7. True. Donald Trump claimed he was using hydroxychloroquine as a preventative treatment for COVID-19, even though studies have suggested that the drug is not effective for this purpose. Source: A. V. Hernandez; et al., 2021. \\
\cmidrule(lr){2-3}
& 8. Cannabis has not been associated with deaths and has potential medical benefits.
& 8. Partly True. The claim is somewhat true as it reflects common perception regarding cannabis safety and therapeutic use, but it oversimplifies a complex topic. Source: D. C. Vidot; et al., 2020. \\

\end{longtable}
\end{singlespace}

\twocolumn 

\end{document}